# Simultaneous two-color snapshot view on ultrafast charge and spin dynamics in a Fe-Cu-Ni tri-layer

Benedikt Rösner,[1*] Boris Vodungbo,[2] Valentin Chardonnet,[2] Florian Döring,[1] Vitaliy A. Guzenko,[1] Marcel Hennes,[2] Armin Kleibert,[1] Maxime Lebugle,[1] Jan Lüning,[2] Nicola Mahne,[3] Aladine Merhe,[2] Denys Naumenko,[4] Ivaylo P. Nikolov,[4] Ignacio Lopez-Quintas,[4] Emanuele Pedersoli,[4] Primož R. Ribič,[4,5] Tatiana Savchenko,[1] Benjamin Watts,[1] Marco Zangrando,[3,4] Flavio Capotondi,[4] Christian David[1], Emmanuelle Jal[2*]

[1] Paul Scherrer Institut, 5232 Villigen PSI, Switzerland

[2] Sorbonne Université, CNRS, Laboratoire de Chimie Physique – Matière et Rayonnement, LCPMR, 75005 Paris, France

[3] IOM-CNR, Strada Statale 14-km 163,5, 34149 Basovizza, Trieste, Italy

[4] Elettra-Sincrotrone Trieste, Strada Statale 14-km 163,5, 34149 Basovizza, Trieste, Italy

[5] Laboratory of Quantum Optics, University of Nova Gorica, 5001 Nova Gorica, Slovenia

**Ultrafast phenomena on a femtosecond timescale are commonly examined by pump-probe experiments. This implies multiple measurements where the sample under investigation is pumped with a short light pulse and then probed with a second pulse at various time delays to follow its dynamics. Recently, the principle of streaking extreme ultraviolet (XUV) pulses in the temporal domain has enabled recording the dynamics of a system within a single pulse. However, separate pump-probe experiments at different absorption edges still lack a unified timing, when comparing the dynamics in complex systems. Here we report on an experiment using a dedicated optical element and the two-color emission of the FERMI XUV free-electron laser to follow the charge and spin dynamics in composite materials at two distinct absorption edges, simultaneously. The sample, consisting of ferromagnetic Fe and Ni layers, separated by a Cu layer, is pumped by an infrared laser and probed by a two-color XUV pulse with photon energies tuned to the M-shell resonances of these two transition metals. The experimental geometry intrinsically avoids any timing uncertainty between the two elements and unambiguously reveals an approximately 100 fs delay of the magnetic response with respect to the electronic excitation for both Fe and Ni. This delay shows that the electronic and spin degrees of freedom are decoupled during the demagnetization process. We furthermore observe that the electronic dynamics of Ni and Fe show pronounced differences when probed at their resonance while the demagnetization dynamics are similar. These observations underline the importance of simultaneous investigation of the temporal response of both charge and spin in multi-component materials. In a more general scenario, the experimental approach can be extended to continuous energy ranges, promising the development of jitter-free transient absorption spectroscopy in the XUV and soft X-ray regimes.**

* corresponding authors: Benedikt.Roesner@psi.ch, emmanuelle.jal@sorbonne-universite.fr

# I. INTRODUCTION

The use of extremely short light pulses is a powerful way to investigate ultrafast phenomena occurring on a femtosecond timescale. With the advent of sources that can provide ultrashort XUV and X-ray pulses, such as femtoslicing, high harmonic generation, and XUV/X-ray free-electron lasers (FELs), the study of ultrafast dynamics specific to an element in condensed matter or atomic and molecular systems is possible in pump-probe experiments[1-3]. In this process, the sample under investigation is excited by a pump pulse and subsequently probed by a second, delayed pulse that is sensitive to the physical effect induced by the excitation. The response of the examined material is followed by varying the time delay between the pump and probe pulses, down to time scales limited by the pulse lengths. This experimental approach, both with optical and XUV/X-ray radiation, provides an opportunity to follow a wide range of ultra-fast processes involving electrons, spins, phonons or other quasi-particles[4-6]. However, some unavoidable difficulties arise in such experiments. For instance, timing inaccuracies are induced by the jitter between pump and probe, e.g. due to synchronization errors between different sources or to mechanical instabilities. Furthermore, the initial state of the investigated system must be fully recovered between the individual exposures to pump and probe pulses. This second aspect limits the potential of ultra-fast, time-resolved experiments severely whenever either the pump or the probe pulses induce irreversible changes, such as static heating that degrades the sample, or the sample recovery time limits the rate at which statistics can be acquired.

These limitations have been overcome using the principle of streaking the arrival time of highly intense XUV FEL pulses at the interaction area. In this specific optical scheme, a whole time trace of a pump-probe experiment is recorded within a single shot from the source[7,8]. This is achieved by stretching the incoming XUV pulse with an off-axis Fresnel zone plate, which introduces an angular encoding of its arrival time at the focus that becomes spatially separated in the far field. The time-streaking method has been demonstrated to provide jitter-free access to a time window of several picoseconds at the M-edges of transition metals[8,9]. So far, the time-streaking method for time-resolved experiments at FELs has been restricted to a monochromatic probe beam. Therefore a single shot dynamics measurement with a unified time scale arising from the same pulse could not be applied to multi-element systems until now. This restriction prevented access to electronic and spin dynamics in complex materials where the excitation can flow through different pathways that involve various species during its time evolution, as available in high harmonic generation (HHG) sources experiments: for instance in charge transfer processes, where the electronic wave-packet migrates to a different atomic site. With HHG sources, several absorption edges can be probed simultaneously at a given time delay, using their intrinsic polychromatic emission. However, time-resolved experiments at HHG sources are still quite time-consuming since the delay between an optical pump pulse and the XUV radiation can only be controlled mechanically. Streaking methods, on the other hand, are still prevented by the intrinsic low number of photons produced by the harmonic conversion process (order of $10^{12}$ - $10^{13}$ photons/sec)[10] compared to a FEL.

An increasing number of technologically relevant systems are being composed of multiple elements with a view towards integrating ultrafast magnetism effects into devices. In such composite systems, the interplay of charge and spin degrees of freedom between the different elements is not yet fully understood[11-14]. For example, tri-layer systems, in which spin currents are generated in an upper ferromagnetic (FM) layer and spread through a non-magnetic (NM) spacer towards a bottom FM layer, have been under intense investigation[15-21]. In a previous study, the observed enhancement of the magnetization of the bottom layer has been attributed to a superdiffusive current created in the top layer[16]. However, this effect has not been observed in other similar systems[17,20], despite a systematic approach changing the NM layer to investigate methods to control this superdiffusive current. In all of these studies, the electronic and magnetization dynamics of both FM layers have been probed separately, and hence information about the relative dynamics between the two magnetic layers has been limited by jitter. The ability to follow the evolution of XUV/X-ray absorption spectra (XAS), including magnetization using X-ray magnetic circular dichroism (XMCD) effects, simultaneously at multiple absorption edges would provide a unified time scale for the set of FM layers and allow new insight into the microscopic dynamics and interplay of such complex systems.

## II. EXPERIMENT

We have developed a custom optical element that allows us to probe the XUV absorption of a composite material simultaneously at both the iron and nickel M-edge within a time window of 2.7 ps (Figure 1). Using the two-color mode of the FERMI FEL tuned at two different harmonics of the seed laser[22], a pair of simultaneous probe pulses is focused onto the sample, and its arrival time is streaked onto the detector with a common time axis that is defined by the fixed geometry. Repeating the measurement with opposing sample magnetizations gives access to the charge and magnetic dynamics. This scheme is applied to study an Fe/Cu/Ni tri-layer in which Fe and Ni are ferromagnetically coupled.

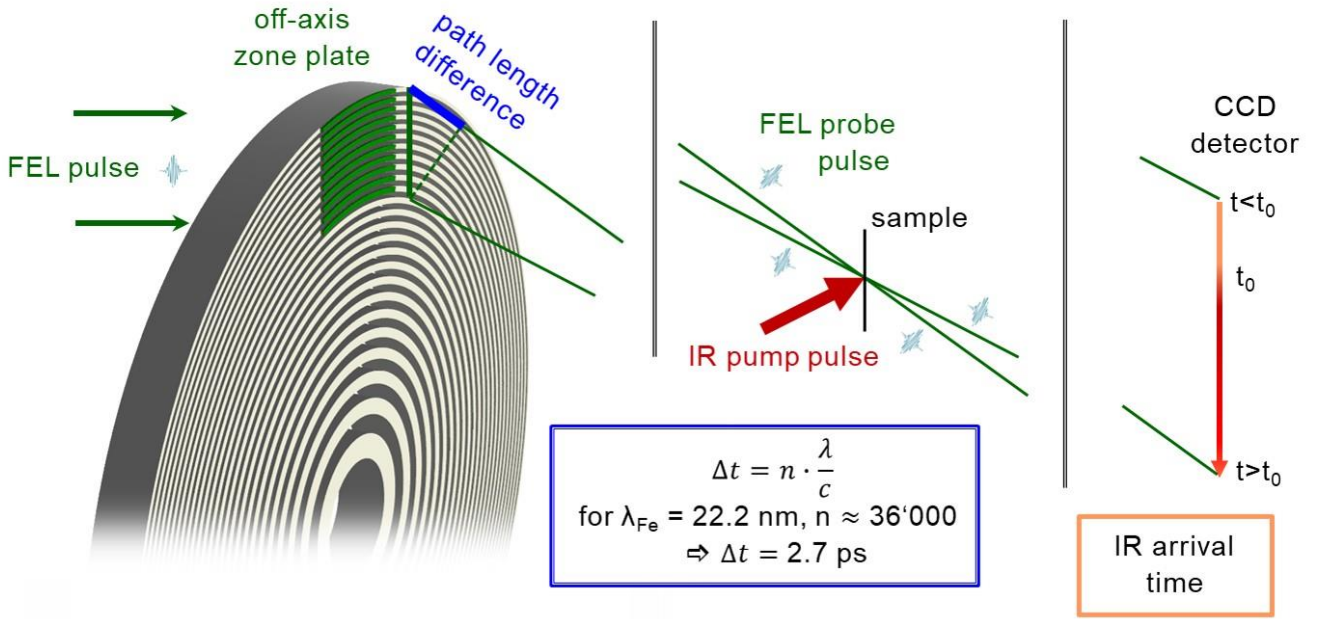

**Figure 1 | Illustration of the time-streaking concept.** An incoming light pulse is diffracted by an off-axis zone plate, which is the outer area of a Fresnel zone plate lens. Inherent to the condition for first order constructive interference, each zone pair induces an exact path difference of the wavelength of the diffracted light. This path length difference then results in a time delay $\Delta t$ that is well defined by the speed of light. In this particular example with 36'000 zone pairs and a wavelength of 22.2 nm, the arrival time of the diffracted beam in the focus spot is stretched over a 2.7 ps time window and projected geometrically (i.e. streaked) onto a two-dimensional detector. By pumping the sample at the focus position, 2.7 ps of its dynamics can be recorded by a single light pulse from a free electron laser.

In previous experiments[8,9], we focused the FEL beam with an off-axis zone plate to a spot in the sample plane that was also optically pumped (Figure 1 and Figure 2a). However, zone plates have a focal length proportional to the photon energy of the incident light and so a single zone plate cannot focus a two-color beam onto a single spot. We therefore designed the two-color experiment to utilize two adjacent off-axis zone plates that share a common optical axis (i.e. two halves of a single zone pattern) and with zone placement designed to focus their respective photon energies to the same focal plane (Figure 2c).

Considering the same displacement from the optical axis and numerical aperture for both cases, two zone plate patterns can be designed for a known energy or wavelength ratio according to:

$$\frac{\Delta r_1}{\Delta r_2} = \frac{E_2}{E_1} = \frac{\lambda_1}{\lambda_2}, \quad \text{(eq. 1)}$$

where $\Delta r$ is the width of the outermost zone.

In order to probe the M-edges of Fe and Ni simultaneously, the electron bunch of the FERMI FEL source was seeded with a single optical laser pulse ($\lambda_{seed} = 245$ nm) and the undulator section was split into two subsections resonant at $\lambda_1 = \lambda_{seed}/m_1$ and $\lambda_2 = \lambda_{seed}/m_2$, where $m_1$ and $m_2$ are integer harmonic numbers. We chose the 13$^{th}$ harmonic to access the Ni M-edge at 65.8 eV ($\lambda_{Ni} = 18.8$ nm), and the 11$^{th}$ harmonic for the Fe M-edge at 55.7 eV ($\lambda_{Fe} = 22.2$ nm). According to eq. (1) and a lower limit of 40 nm arising from the nanofabrication process, the ratio of the outermost zone width is given by:

$$\frac{\Delta r_{Fe}}{\Delta r_{Ni}} = \frac{\lambda_{Fe}}{\lambda_{Ni}} = \frac{m_{Ni}}{m_{Fe}} = \frac{13}{11} = \frac{47.3 \text{ nm}}{40.0 \text{ nm}} \quad \text{(eq. 2)}$$

Since a binary zone plate does not cause wavelength-dependent phase retardation (and hence different optical path lengths for the two colors), both beams travel through path lengths determined by geometry to be exactly the same distance. Thus, the time window is identical for both colors. Note that each half of the off-axis zone plate will diffract both beam components and will therefore produce two additional first order focus spots with longer and shorter focal length than the desired overlapping focal plane. However, these parasitic beams are well separated from the two-color focus and can be spatially filtered by the aperture formed by the frame of the sample support.

The previously developed streaking approach recorded the negative, divergent diffraction order on a second camera for an intensity normalization to eliminate the effects of the significant shot-to-shot fluctuations inherent to the incident FEL beam[8,9]. However, this intensity normalization scheme cannot be extended to the multi-color case since the negative diffraction order of two different wavelengths cannot be separated in space on the reference detector. To address this issue, we combined our two-part, off-axis zone plate with a phase grating that splits the beam perpendicular to the diffraction angle of the off-axis zone plate. An elegant way to integrate a beam splitter in a diffractive optical element is a periodic inversion of the zone pattern as shown in Figure 2 b), resulting in a twin spot of the off-axis zone plate in the focal plane with a defined separation. In this way, one of these spots can be optically pumped while the other records the sample absorption in the unperturbed state. A consequence of this focus splitting is that the beam paths followed by the two colors is no longer exactly equal, but is slightly shorter for the color coming from the off-axis zone plate half on the same side as the focus. This path length difference presents itself in the detector image as a tilt of the line describing equal time values since the path length difference is minimal at the boundary of the color sections and increases (in the corresponding sense) linearly with the distance from the color boundary (Figure 2c).

The periodicity $p$ of the zone inversion for the beam splitter is chosen in the same ratio as the outermost zone widths in order to achieve the same focus spot displacement for the two colors:

$$\frac{p_{Fe}}{p_{Ni}} = \frac{13}{11} = \frac{709 \text{ nm}}{600 \text{ nm}} \quad \text{(eq. 3)}$$

With this set of parameters, we designed off-axis zone plates with an area of 3.8 mm x 3.8 mm, which is part of a virtual zone plate lens with a diameter of $D = 35.0$ mm. The resulting focal length is 74.4 mm, the spot displacement from the optical axis is 2.3 mm. The optical elements were fabricated by etching nanostructures into a 200 nm thick silicon membrane purchased from Norcada, Inc. using a process described elsewhere[23,24].

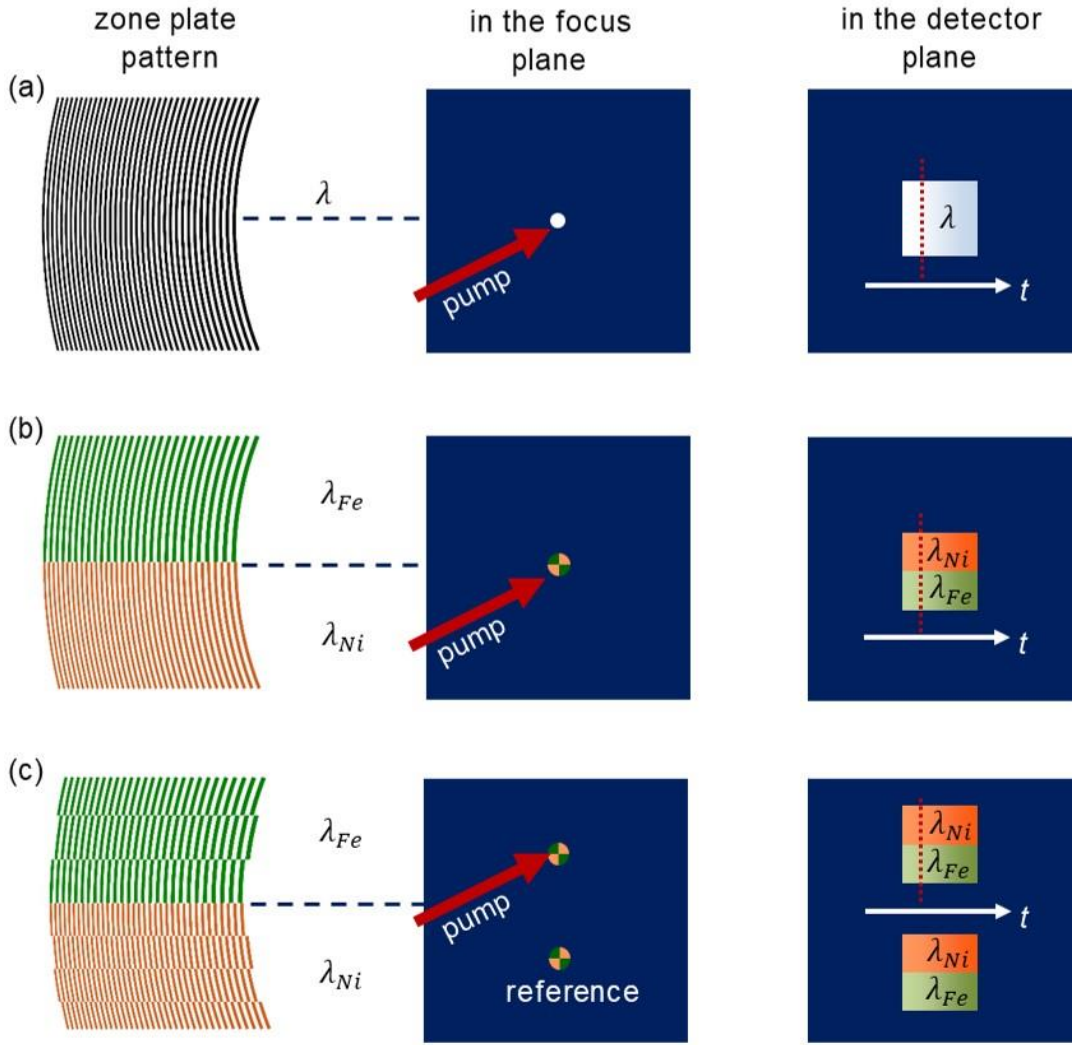

**Figure 2 | Twin focus off-axis zone plate for two colors. a,** Off-axis zone plate as previously used for time-streaking experiments[8,9]. **b,** Two off-axis zone plate patterns are placed next to each other in a way that a two-color beam with two compound energies is focused onto the same spot. **c,** A beam splitter grating perpendicular to the zones is integrated by periodic inversion of the zone plate patterns. This splits the focus spot into twin copies with defined separation, and introduces a slight tilt to the locus of equal time points in the detector image.

Figure 3a shows the recorded image of a shadow mask test pattern placed upstream of the two-color, twin-focus, off-axis zone plate and attests to its quality by the homogeneous, magnified double-projection of the test pattern onto the detector. Each replica of the shadow mask projection consists of two well-defined halves, which can be attributed to the two halves of the zone plate pattern and their corresponding photon energies. Note that the intensity of the FEL beam at 55.7 eV ($\lambda_{Fe}$ = 22.2 nm) is lower compared to the radiation with higher photon energy at 65.8 eV ($\lambda_{Ni}$ = 18.8 nm), and follows the image brightness in the upper and lower halves of each test pattern image. This is due to the energy-dependent beamline transmission, and more particularly due to the absorption of the 200 nm Al solid-state attenuators that were used to protect the sample from damage. Looking at the edges of the depicted logos, we can see that the lower and upper halves are exactly aligned, indicating that our imaging optical element does not introduce artifact into the object lateral dimension and thus, the timing of the two beams matches. As illustrated in Figure 2c, the projection below the optical axis is used as a reference, while the upper one is used to track the pump-induced absorption changes. Division of one image by the other gives a precise normalization of the incident two-color beam on a shot-to-shot basis (see supplementary material for details).

Using this set up, we investigated a magnetic sample that consists of a tri-layer of polycrystalline metallic films (10 nm Ni, 2 nm Cu, and 10 nm Fe), all grown on top of a 3 nm thin metallic Ta adhesion layer by sputtering. The Ta adhesion layer was deposited directly onto silicon nitride ($Si_3N_4$) membranes (200 x 200 µm$^2$) of 30 nm thickness. To prevent oxidation, the Ni film was capped with a 3 nm thin Al layer. The detailed structure of the entire stack, $Si_3N_4$(30)/Ta(3)/Fe(10)/Cu(2)/Ni(10)/Al(3) – the number in brackets is in nanometers –, is shown in Figure 4a. Static magneto-optic Kerr effect measurements confirmed the expected in-plane magnetization of the Ni and Fe films, which are ferromagnetically coupled through the Cu layer. The tri-layer sample was placed in the focal plane of the twin-focus, off-axis zone plates such that one membrane window was brought to the upper focus spot and pumped by the IR laser (780 nm wavelength, 100 fs pulse duration, 2.5 µJ incident pulse energy, and a spot size of 260x140 µm$^2$). Note that both the Ni and the Fe layer are similarly excited at the used pump fluence (see supplementary information). Another part of the magnetic layer system was aligned such that the lower focus spot passed through it to obtain an unperturbed reference beam. In this way, the two projections on the detector contain information on the transmitted signal from a pumped sample region, as well as the unpumped reference as described in the previous paragraph. The ratio of these two images yields the transient transmission of the sample when it is excited. Repeating the measurement with a magnetic field of approximately 130 mT applied in opposite direction allows us to retrieve XAS and XMCD signals from the sum and difference of the measurements, respectively. While XAS is sensitive to the transient changes in the electronic system, XMCD is a direct probe of the spin state of the magnetic system. For probing the in-plane magnetization, the sample was tilted by 18° with respect to the incident beam. More information on the experimental geometry can be found in the supplemental material.

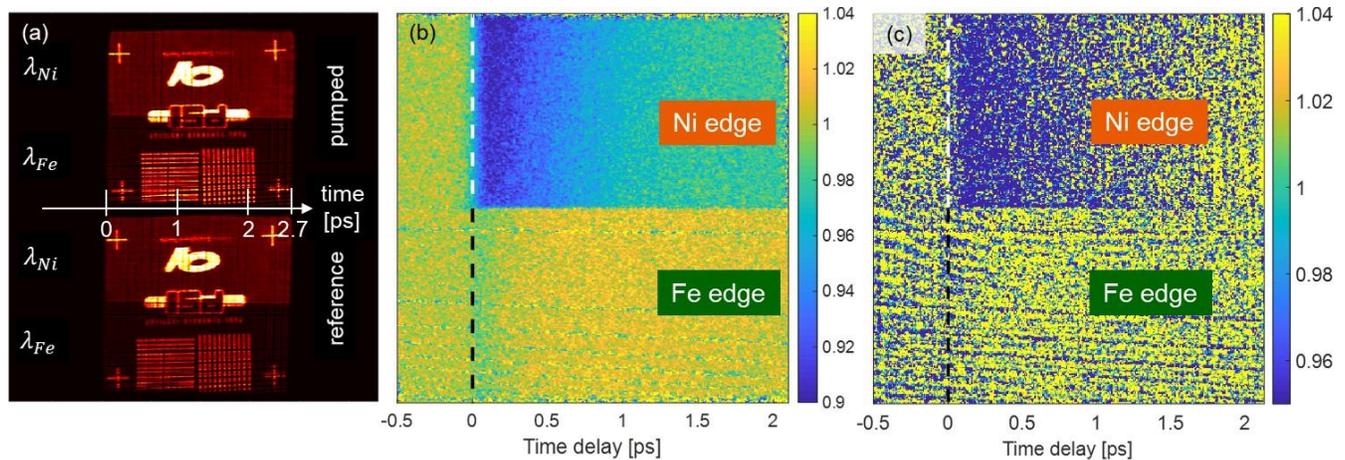

**Figure 3 | Time-streaking experiment at two absorption edges. a,** Projection of a shadow mask placed in front of the zone plate onto the detector. The two-color beams are visible as a distinct contrast on the recorded image due to different pulse intensity. The arrival time of the 2-color probe beams is dispersed along the horizontal direction. In the time resolved experiment, one focus spot is pumped with an IR laser while the unpumped one is used as reference. Note that the detector image has been rotated by 180° to the reader's perspective. **b,** Temporal evolution of the transmission of a pumped iron-nickel tri-layer sample after optical excitation (the dashed line identifies t = 0) integrated over 600 shots. The upper part corresponds to the Ni M-edge, while the lower part shows the time evolution of the signal at the Fe M-edge. The reported images are normalized and corrected with respect to the zone plate curvature using the same procedure as described by Buzzi *et al.*, and Jal *et al.*[8,9], and with respect to the time delay induced by the beam splitter grating. **c,** Respective image recorded with a single FEL pulse exposure.

In order to account for variations in the local absorption and detector response, the normalized picture has also to be divided by background images taken without the infrared pump. Since the time delay is a function of the zone placement, the curvature of the zone plate has to be corrected as reported before[8,9]. The resulting picture retrieved for an XUV fluence of 1 mJ/cm$^2$ and an accumulation of 600 shots is shown in Figure 3b. We would like to emphasize here that it is possible to measure a signal even with one single shot, as shown in Figure 3c (XUV fluence of 4 mJ/cm$^2$). We will, however, concentrate on the data taken for an accumulation of 600 shots in the following as the process we are investigating is fully reversible.

## III. RESULTS AND DISCUSSION

Integration of the two areas in Figure 3b with the different photon energies shows the evolution of XUV transmission in time. The logarithm of this normalized signal directly gives the relative absorption $\Delta\mu^+(t)$ and $\Delta\mu^-(t)$ for the two opposite magnetization directions, corresponding to the difference between the dynamic and static absorption ($\Delta\mu^{+/-}(t) = \Delta\mu^{+/-}(t) - \Delta\mu^{+/-}(t<0)$, see supplementary material). The average of these two signals shows the time evolution of the absorption, $\Delta$XAS, and the difference gives the change in magnetic contrast with time, $\Delta$XMCD. $\Delta$XAS is sensitive to the transient electronic population whereas $\Delta$XMCD is proportional to the transient electronic spin dynamics[25]. In order to compare both elements, we have to normalize the relative $\Delta$XAS and $\Delta$XMCD by the static XAS$_0$ and XMCD$_0$. Figure 4b shows the $\Delta\mu^{+/-}(t)/\mu_0^{+/-}$ response of the Ni (red) and Fe (blue) layers, respectively. If we zoom around *t$_0$* (Figure 4c), we can clearly observe that the absorption for both applied magnetic fields is equal within the first 100 fs, meaning that the spin dynamics for both elements are delayed by approx. 100 fs with respect to the electronic response (grey area in Figure 4c). Moreover, the electronic temporal dependence of the Fe film is strikingly different from that of the Ni film, while the magnetic behavior is similar, as demonstrated by the comparison of $\Delta$XAS/XAS$_0$ and $\Delta$XMCD/XMCD$_0$ of both elements in Figure 4d. While Figure 4c and 4e highlight a delay of the XMCD signal with respect to the electronic response, we can clearly observe in Figure 4d and 4e that there is no relative delay between the two species (Fe and Ni). Together with the very recent study of Yao *et al.*[26], we are among the first reporting such a delay in the spin degree of freedom with respect to an excitation in the electronic system for ferromagnetic elements, although a delayed response of spin dynamics has been predicted theoretically by Zhang *et al.*[27,28]. Other measurements performed on a completely different system (that will be presented and discussed in a future publication), as well as the work of Yao *et al.* mentioned above, seem to indicate that the observed phenomenology is specific to the M-edge excitation and depends on the probing photon energies. In more detail, Yao *et al.* explain their observations by the competition, at specific photon energies and on short time scales, of two competitive transient contributions with opposite sign. The first one is a red shift of the absorption edge that increases the magnetic scattering cross section for a photon energy below the resonance condition, and the second one is an ultrafast demagnetization that, on the other hand, reduces the XMCD signal. This would explain why earlier time-resolved XAS and XMCD studies, which have been performed at the L$_3$-edge of the 3d transition metals, have not observed a delayed onset of the electronic and magnetic dynamics[29,30].

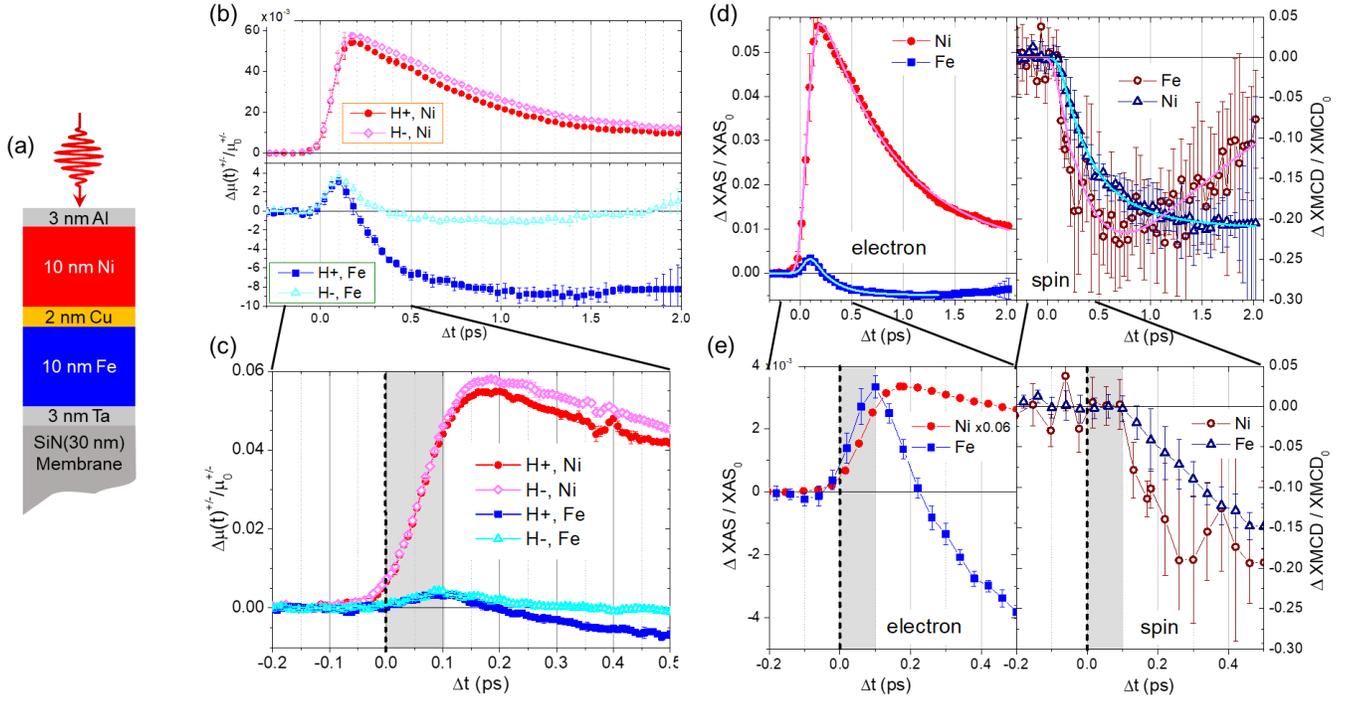

**Figure 4 | Jitter-free electronic and spin dynamics of iron and nickel probed simultaneously with the same absolute timing. a,** Layer structure of the investigated tri-layer film. **b,** Time traces of the relative absorption for nickel (red, upper part) and iron (blue, lower part) for opposite magnetic field direction + and -. **d,** The average signal of the traces recorded at opposite magnetic field shown in (b) gives the relative XUV absorption $\Delta XAS/XAS_0$ and thus a measure for the electronic dynamics (solid symbols, left panel), whereas the difference of these two signals gives the $\Delta XMCD/XMCD_0$ signal that represents the spin dynamics (open symbols, right panel). **c and e,** zoom of Figure (b) and (d) respectively, where the grey area represents the 100 fs onset difference between electronic and spin signal and $t_0$ is marked with a dashed line. **c,** the relative absorption for + and – magnetic field direction separates from each other around 100 fs. Note that the error bars of (b) are coming from the averaging over 5 points of curves in (c) where one point correspond to one pixel in the x direction. **d,** The direct comparison of $\Delta XAS/XAS_0$ and $\Delta XMCD/XMCD_0$ for Ni and Fe shows that the electronic dynamics (left panel) begin sooner than the spin dynamics (right hand side). Note that the $\Delta XAS/XAS_0$ curve for Ni has been scaled down for better comparability.

Although this delay between $\Delta XMCD/XMCD_0$ and $\Delta XAS/XAS_0$ has not been observed in previous experiments at the $L_3$-edge, we emphasize that the marked dynamics we measured here for $\Delta XAS/XAS_0$ of Ni are very similar to the one measured by Stamm *et al.*[29] and predicted by Carva *et al.*[31] at the $L_3$-edge of Ni. According to the latter work, these dynamics arise from a change of the electron density of states induced by the IR pump. Since the M-edge is wider than the $L_3$-edges and as we are not probing on the flat top part of the Ni absorption edge ($\approx 67$ eV) but at slightly lower photon energy on the steep slope of the absorption spectrum (Figure 5a), we are very sensitive to the change of electronic density of states in the conduction band. In our case, most of the IR photons are absorbed by valence electrons during the excitation of the sample, and Ni valence electrons that lie 1.5 eV below the Fermi level are promoted into empty states above it. As pointed out by Carva *et al.* in their simulations[31], this process will generate hot electrons[32] that cannot be modeled by a Fermi-Dirac distribution, and will produce a dynamical shift of the XAS spectrum towards lower energies, opening new available states to perform the electronic transition from the core level to the conduction band. This will produce an increase of the sample absorption for photons below the resonance energy on ultrashort time scales, as illustrated by the red rectangle in Figure 5a. On longer time scales, the excited electrons relax and fill the empty states in the valence band, thereby reducing the number of available transitions for the XUV absorption events at photon energies below the resonance condition. In order to extract characteristic times from our data, all curves have been fitted by a two exponential model[15] to obtain the typical time constants. The fits are shown in Figure 4d, and the fit parameters are given in the supplemental material and discussed in the next paragraphs. For $\Delta XAS/XAS_0$ of Ni, the double exponential fit shown in Figure 4d yields a rise time constant of 60 fs with a decay time constant of 0.8 ps. This rise time constant is in good agreement with the lifetime of optically excited electrons in Ni[33-35].

If we apply the same reasoning to Fe where the probing energy is closer to the absorption peak (55.7 eV, Figure 5a), the 1.5 eV excitation will lead to an initial increase of the transient XUV absorption and a subsequent decrease already underway before the onset of the slower relaxation process (blue rectangle in Figure 5a). These dynamics are indeed observed in our experiments as demonstrated in Figure 4d and 4e. The double exponential fit in Figure 4d yields a rising time constant of 20 fs for Fe, and a time constant of 200 fs for its decrease. Note that in our streaking approach the contribution of the optical geometry to the temporal resolution is as low as 2.5 fs and can be neglected here.

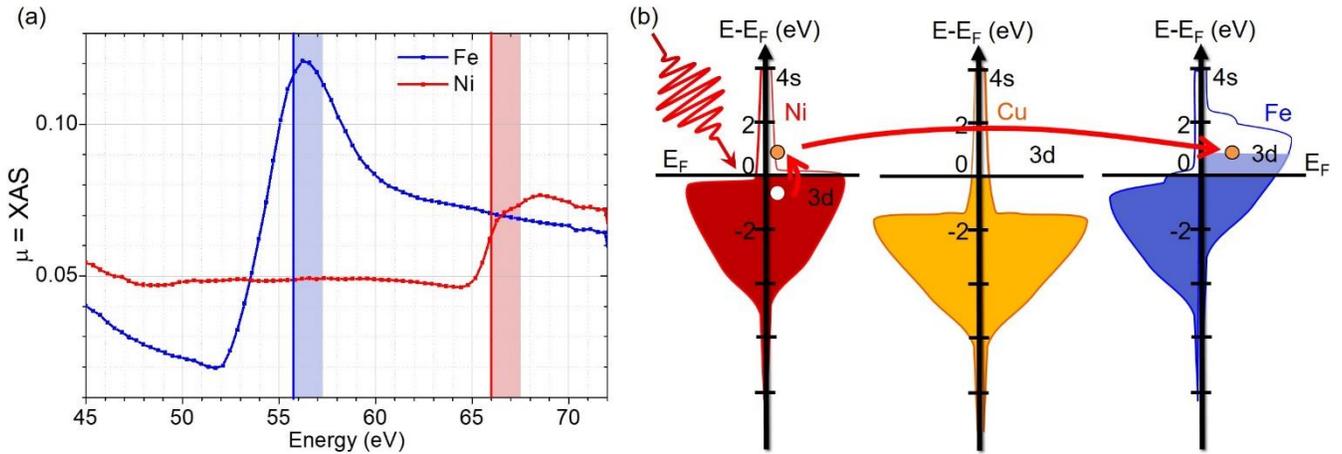

**Figure 5 | Microscopic model for the observed data. a,** Static absorption as reported by Willems *et al.*[22]. The vertical lines indicate the probed energies while the shaded areas indicate the 1.5 eV excitation by the IR laser. **b,** Simplified density of states of Ni, Cu and Fe, adapted from Zhukov *et al.*[33,39]. The empty states in Ni are mostly itinerant s-states whereas mostly localized d-states are present in Fe. This implies that excited Ni electrons are mainly itinerant and can travel through the Cu layer (red arrow) to fill the empty states of the Fe minority band (light blue area), reducing the Fe magnetization further.

At first glance, the Ni and Fe subsystems display a classical ultrafast demagnetization behavior. In Figure 4d we observe that both Fe and Ni are demagnetized by 20%. However, the Ni layer demagnetized faster than the Fe layer with a demagnetization time constant of 250 fs while the Fe layer has a constant of 450 fs. Compared to previous results in the literature[29,36,37], our data shows a demagnetization that is slightly slower for Ni and more than twice as slow as usually observed for Fe. This could be explained by the fact that we are not probing exactly at the element-specific resonances, which are at 54.2 eV for Fe and 66.2 eV for Ni, but 1.5 eV above the Fe M-edge and 0.4 eV below the Ni M-edge. Indeed, a recent paper by Gort *et al.*[38] points out that the observed dynamical behavior of the spin polarization depends not only on time but also on the binding energy of the probed electrons with respect to the Fermi level. Furthermore, the slower demagnetization time for Fe could be due to a possible pure spin current travelling between Ni and Fe at a longer time scale [40,41]. Within a simple band model as shown in Figure 5b, a unidirectional transfer of Ni minority electrons into the Fe minority band can be possible. If we consider the band structure of Fe and Ni, there are mainly d-states available up to 1.5 eV above the Fermi level in Fe, while Ni possesses only s-states (Figure 5b). This implies that the excited electrons are almost completely in the localized 3d band in Fe, while they are in the itinerant 4s band in Ni. Thus, excited electrons in Ni can flow across the Cu layer, and fill the empty states of the minority band of Fe, leading to a decrease in the overall magnetization of Fe in addition to the intrinsic laser-induced demagnetization. Although the limited thickness of our tri-layer system implies that all magnetic layers are optically excited (the Ni layer absorbs 10.6% of the incoming IR fluence while the Fe layer absorbs 12.2%, see supplementary material for the complete IR absorption profile), it seems that there is no transport of the Fe excited electrons into the Ni. This is mainly because the excited Fe electrons are localized in the d-states available up to 1.5 eV above the Fermi level.

Additionally, we would like to point out the difference between the spin transport mechanism proposed above and the optically induced spin and orbital momentum transfer occurring in the first few femtoseconds, described recently by Siegrist *et al.*[41]. In our case, the spin transport occurs on longer timescales and over more than 2 nm of Cu. We therefore hope our results will encourage new theoretical simulations to understand the interplay of electron dynamics in coupled 3d metals over a time range of several hundreds of femtoseconds in detail.

Finally, like in the previous work of Turgut *et al*[17], and Alekin *et al*[20], for a tri-layer with a non-magnetic layer different than Ru, we do not observe any enhancement of the magnetization due to superdiffusive currents. We do not exclude any superdiffusive current mechanism in the demagnetization process, without more systematic measurements on the same kind of multilayer exhibiting anti-parallel magnetic coupling between the Fe and Ni layers. But, we would like to point out that our pure absorption technique is directly sensitive to the magnetic moment of Ni and Fe, while in T-MOKE experiments, the complexity of the magneto-optical response in the XUV regime has recently been highlighted[42]. This study shows strong nonlinearity in the magnetic asymmetry of Ni and Fe transverse magneto optical Kerr effect in the energy range between 40 to 70 eV. This strong dependence of the magnetic asymmetry has been explained as a direct consequence of the different type of magnetic excitations involved into the fast demagnetization process for the two material, i.e. longitudinal Stoner excitations or transversal short or long wavelengths magnon emission. This could explain why in their recent study Stamm *et al.* do also not observe an enhancement of the transient magnetization of the Fe layer in a Ni/Ru/Fe sandwich[43]. It also highlights the importance to perform more systematic time resolved studies in the pure absorption geometry where the measured intensities can be more easily related to physical concepts such as electronic density and magnetization.

## IV. CONCLUSION

Our results demonstrate that it is possible to investigate the dynamics of a multicomponent system at two discrete photon energies with absolute timing. The special benefit of this method arises from the new normalization scheme based on a beam splitter grating that is directly integrated into the off-axis zone plate used for the time-streaking method[8,9]. In this way, it is not only possible to conduct an entire pump-probe experiment in a single snapshot, but also to design parts of the zone plate for discrete energies in a fixed geometry, and thus to maintain a uniform time scale across the dynamics of multiple elements. While the very good, jitter-free temporal resolution is an inherent advantage of the streaking approach[9], the absolute temporal comparison of the starting dynamics for both Fe and Ni elements is only possible when using a two-color pulse to probe the system. This is the motivation for the development of the two-color schemes described in this paper. Note that this technique cannot be applied at HHG sources, which do not provide enough flux to be operated with optics that have limited efficiency. To the best of our knowledge, this is the first time that a single snapshot of delay traces has been combined with simultaneous probing of several energies in the XUV regime.

By investigation of magnetism in the femtosecond regime, this study shows the potential, and the necessity, of simultaneously probing spin and charge in heterostructures. This multi-parameter approach leads to valuable new insights into the existing theory and to a more comprehensive understanding of the microscopic mechanisms that are responsible for ultrafast electronic processes. Our two-color XUV streaking experiment has, for instance, the potential to resolve the controversy of small time delays in the onset of demagnetization of Fe and Ni sublattices[36,37,44] in alloys.

Taking a step further, the extension of this method to continuous energies opens up perspectives for single-shot transient spectroscopy in the XUV and soft X-ray regime, in particular at facilities that offer ultrashort and large bandwidth pulses. This paves the way towards real femtosecond single-shot transient spectroscopy conducted at X-ray free electron laser facilities.

## SUPPLEMENTARY MATERIAL

See supplementary material for more information about the experimental geometry, the normalization and details to retrieve XAS and XMCD signals, as well as for the fitting parameters and absorption profile.

## ACKNOWLEDGEMENTS

This work was funded within the EU-H2020 Research and Innovation Programme, No. 654360 NFFA-Europe (BR). The authors are grateful for financial support received from the CNRS-MOMENTUM, the SNSF project (No. 200021_160186), the UMAMI ANR-15-CE24-0009, and the CNRS-PICS programs. Access to Synchrotron SOLEIL and beamline SEXTANTS through proposal ID 20160880 for characterization of static properties of the tri-layers is acknowledged. We also thank Clemens von Korff Schmising and Peter Oppeneer for stimulating discussion.

## DATA AVAILABILITY

The data that support the findings of this study are openly available under the Creative Common license in the Zenodo repository, https://doi.org/10.5281/zenodo.3735723[45].

## REFERENCES

1  M. Beye, F. Sorgenfrei, W. F. Schlotter, W. Wurth and A. Föhlisch. P. Natl. Acad. Sci. USA 107, 16772 (2010), https://doi.org/10.1073/pnas.1006499107.
2  R. J. Squibb, M. Sapunar, A. Ponzi, R. Richter, A. Kivimäki, O. Plekan, P. Finetti, N. Sisourat, V. Zhaunerchyk, T. Marchenko *et al.* Nat. Commun. 9, 63 (2018), https://doi.org/10.1038/s41467-017-02478-0.
3  C. L. Smallwood, J. P. Hinton, C. Jozwiak, W. Zhang, J. D. Koralek, H. Eisaki, D.-H. Lee, J. Orenstein and A. Lanzara. Science 336, 1137 (2012), https://doi.org/10.1126/science.1217423.
4  E. Beaurepaire, J. C. Merle, A. Daunois and J. Y. Bigot. Phys. Rev. Lett. 76, 4250-4253 (1996), https://doi.org/10.1103/PhysRevLett.76.4250.
5  F. Bencivenga, R. Cucini, F. Capotondi, A. Battistoni, R. Mincigrucci, E. Giangrisostomi, A. Gessini, M. Manfredda, I. P. Nikolov, E. Pedersoli *et al.* Nature 520, 205 (2015), https://doi.org/10.1038/nature14341.
6  D. Fausti, R. I. Tobey, N. Dean, S. Kaiser, A. Dienst, M. C. Hoffmann, S. Pyon, T. Takayama, H. Takagi and A. Cavalleri. Science 331, 189 (2011), https://doi.org/10.1126/science.1197294.
7  C. David, P. Karvinen, M. Sikorski, S. Song, I. Vartiainen, C. J. Milne, A. Mozzanica, Y. Kayser, A. Diaz, I. Mohacsi *et al.* Sci. Rep. 5, 7644 (2015), http://dx.doi.org/10.1038/srep07644.

8. M. Buzzi, M. Makita, L. Howald, A. Kleibert, B. Vodungbo, P. Maldonado, J. Raabe, N. Jaouen, H. Redlin, K. Tiedtke et al. Sci. Rep. 7, 7253 (2017), https://dx.doi.org/10.1038/s41598-017-07069-z.
9. E. Jal, M. Makita, B. Rösner, C. David, F. Nolting, J. Raabe, T. Savchenko, A. Kleibert, F. Capotondi, E. Pedersoli et al. Phys. Rev. B 99, 144305 (2019), https://doi.org/10.1103/PhysRevB.99.144305.
10. S. Hädrich, A. Klenke, J. Rothhardt, M. Krebs, A. Hoffmann, O. Pronin, V. Pervak, J. Limpert and A. Tünnermann. Nat. Photon. 8, 779-783 (2014), https://doi.org/10.1038/nphoton.2014.214.
11. C. E. Graves, A. H. Reid, T. Wang, B. Wu, S. de Jong, K. Vahaplar, I. Radu, D. P. Bernstein, M. Messerschmidt, L. Müller et al. Nat. Mater. 12, 293 (2013), https://doi.org/10.1038/nmat3597.
12. E. Iacocca, T. M. Liu, A. H. Reid, Z. Fu, S. Ruta, P. W. Granitzka, E. Jal, S. Bonetti, A. X. Gray, C. E. Graves et al. Nat. Commun. 10, 1756 (2019), https://doi.org/10.1038/s41467-019-09577-0.
13. P. W. Granitzka, E. Jal, L. Le Guyader, M. Savoini, D. J. Higley, T. Liu, Z. Chen, T. Chase, H. Ohldag, G. L. Dakovski et al. Nano Lett. 17, 2426-2432 (2017), https://doi.org/10.1021/acs.nanolett.7b00052.
14. E. Ferrari, C. Spezzani, F. Fortuna, R. Delaunay, F. Vidal, I. Nikolov, P. Cinquegrana, B. Diviacco, D. Gauthier, G. Penco et al. Nat. Commun. 7, 10343 (2016), https://doi.org/10.1038/ncomms10343.
15. G. Malinowski, F. Dalla Longa, J. H. H. Rietjens, P. V. Paluskar, R. Huijink, H. J. M. Swagten and B. Koopmans. Nat. Phys. 4, 855-858 (2008), https://doi.org/10.1038/nphys1092.
16. D. Rudolf, C. La-O-Vorakiat, M. Battiato, R. Adam, J. M. Shaw, E. Turgut, P. Maldonado, S. Mathias, P. Grychtol, H. T. Nembach et al. Nature Commun. 3, 1037 (2012), https://doi.org/10.1038/ncomms2029.
17. E. Turgut, C. La-o-vorakiat, J. M. Shaw, P. Grychtol, H. T. Nembach, D. Rudolf, R. Adam, M. Aeschlimann, C. M. Schneider, T. J. Silva et al. Phys. Rev. Lett. 110, 197201 (2013), https://doi.org/10.1103/PhysRevLett.110.197201.
18. A. J. Schellekens, N. de Vries, J. Lucassen and B. Koopmans. Phys. Rev. B 90, 104429 (2014), https://doi.org/10.1103/PhysRevB.90.104429.
19. G.-M. Choi, B.-C. Min, K.-J. Lee and D. G. Cahill. Nat. Commun. 5, 4334 (2014), https://doi.org/10.1038/ncomms5334.
20. A. Alekhin, I. Razdolski, N. Ilin, J. P. Meyburg, D. Diesing, V. Roddatis, I. Rungger, M. Stamenova, S. Sanvito, U. Bovensiepen et al. Phys. Rev. Lett. 119, 017202 (2017), https://doi.org/10.1103/PhysRevLett.119.017202.
21. A. Eschenlohr, L. Persichetti, T. Kachel, M. Gabureac, P. Gambardella and C. Stamm. J. Phys. Condens. Matter 29, 384002 (2017), https://doi.org/10.1088/1361-648X/aa7dd3.
22. F. Willems, C. von Korff Schmising, D. Weder, C. M. Günther, M. Schneider, B. Pfau, S. Meise, E. Guehrs, J. Geilhufe, A. E. D. Merhe et al. Struct. Dynam. 4, 014301 (2017), https://doi.org/10.1063/1.4976004.
23. B. Rösner, F. Döring, P. R. Ribič, D. Gauthier, E. Principi, C. Masciovecchio, M. Zangrando, J. Vila-Comamala, G. De Ninno and C. David. Opt. Express 25, 30686-30695 (2017), https://doi.org/10.1364/OE.25.030686.
24. P. R. Ribič, B. Rösner, D. Gauthier, E. Allaria, F. Döring, L. Foglia, L. Giannessi, N. Mahne, M. Manfredda, C. Masciovecchio et al. Phys. Rev. X 7, 031036 (2017), https://doi.org/10.1103/PhysRevX.7.031036.
25. J. Stöhr and H. C. Siegmann. *Magnetism*. Vol. 152 (Springer-Verlag, 2006).

26 K. Yao, F. Willems, C. von Korff Schmising, I. Radu, C. Strueber, D. Schick, D. Engel, A. Tsukamoto, J. K. Dewhurst, S. Sharma, S. Eisebitt arXiv 2005.02872 (2020), https://arxiv.org/abs/2005.02872.
27 G. P. Zhang, Y. H. Bai, T. Jenkins and T. F. George. J. Phys. Condens. Matter 30, 465801 (2018), https://doi.org/10.1088/1361-648x/aae5a9.
28 G. P. Zhang, W. Hübner, G. Lefkidis, Y. Bai and T. F. George. Nat. Phys. 5, 499-502 (2009), https://doi.org/10.1038/nphys1315.
29 C. Stamm, T. Kachel, N. Pontius, R. Mitzner, T. Quast, K. Holldack, S. Khan, C. Lupulescu, E. F. Aziz, M. Wietstruk *et al.* Nat. Mater. 6, 740-743 (2007), https://doi.org/10.1038/nmat1985.
30 C. Boeglin, E. Beaurepaire, V. Halté, V. López-Flores, C. Stamm, N. Pontius, H. A. Dürr and J. Y. Bigot. Nature 465, 458-461 (2010), https://doi.org/10.1038/nature09070.
31 K. Carva, D. Legut and P. M. Oppeneer. Europhys. Lett. 86, 57002 (2009), https://doi.org/10.1209/0295-5075/86/57002.
32 S.I. Anisimov, B. L. Kapeliovich, and T. L. Perel'man. Sov. Phys. JETP. 39, 375 (1974), http://www.jetp.ac.ru/cgi-bin/e/index/e/39/2/p375?a=list.
33 V. P. Zhukov, E. V. Chulkov and P. M. Echenique. Phys. Rev. Lett. 93, 096401 (2004), https://doi.org/10.1103/PhysRevLett.93.096401.
34 V. P. Zhukov, E. V. Chulkov and P. M. Echenique. Phys. Rev. B 73, 125105 (2006), https://doi.org/10.1103/PhysRevB.73.125105.
35 M. Bauer, A. Marienfeld and M. Aeschlimann. Prog. Surf. Sci. 90, 319-376 (2015), https://doi.org/10.1016/j.progsurf.2015.05.001.
36 C. La-O-Vorakiat, E. Turgut, C. A. Teale, H. C. Kapteyn, M. M. Murnane, S. Mathias, M. Aeschlimann, C. M. Schneider, J. M. Shaw, H. T. Nembach *et al.* Phys. Rev. X 2, 011005 (2012), https://doi.org/10.1103/PhysRevX.2.011005.
37 S. Mathias, C. La-O-Vorakiat, P. Grychtol, P. Granitzka, E. Turgut, J. M. Shaw, R. Adam, H. T. Nembach, M. E. Siemens, S. Eich *et al.* P. Natl. Acad. Sci. USA 109, 4792 (2012), https://doi.org/10.1073/pnas.1201371109.
38 R. Gort, K. Bühlmann, S. Däster, G. Salvatella, N. Hartmann, Y. Zemp, S. Holenstein, C. Stieger, A. Fognini, T. U. Michlmayr *et al.* Phys. Rev. Lett. 121, 087206 (2018), https://doi.org/10.1103/PhysRevLett.121.087206.
39 V. P. Zhukov and E. V. Chulkov. Phys. Usp. 52, 105-136 (2009), https://doi.org/10.3367/ufne.0179.200902a.0113.
40 V. Shokeen, M. Sanchez Piaia, J. Y. Bigot, T. Müller, P. Elliott, J. K. Dewhurst, S. Sharma and E. K. U. Gross. Phys. Rev. Lett. 119, 107203 (2017), https://doi.org/10.1103/PhysRevLett.119.107203.
41 F. Siegrist, J. A. Gessner, M. Ossiander, C. Denker, Y.-P. Chang, M. C. Schröder, A. Guggenmos, Y. Cui, J. Walowski, U. Martens *et al.* Nature 571, 240-244 (2019), https://doi.org/10.1038/s41586-019-1333-x.
42 S. Jana, R. S. Malik, Y. O. Kvashnin, I. L. M. Locht, R. Knut, R. Stefanuik, I. Di Marco, A. N. Yaresko, M. Ahlberg, J. Åkerman, R. Chimata, *et al.* Phys. Rev. Research 2, 013180 (2020), https://doi.org/10.1103/PhysRevResearch.2.013180.
43 C. Stamm, C. Murer, M. S. Wörnle, Y. Acremann, R. Gort, S. Däster, A. H. Reid, D. J. Higley, S. F. Wandel, W. F. Schlotter, and P. Gambardella J. Appl. Phys. 127, 223902 (2020) https://aip.scitation.org/doi/10.1063/5.0006095.
44 I. Radu, C. Stamm, A. Eschenlohr, F. Radu, R. Abrudan, K. Vahaplar, T. Kachel, N. Pontius, R. Mitzner, K. Holldack *et al.* Spin 05, 1550004 (2015), https://doi.org/10.1142/s2010324715500046.

45  B. Rösner, B. Vodungbo, V. Chardonnet, F. Döring, V. A. Guzenko, M. Hennes, A. Kleibert, M. Lebugle, J. Lüning, N. Mahne *et al.*(2020) Simultaneous two-color snapshot view on ultrafast charge and spin dynamics in a Fe-Cu-Ni tri-layer. Zenodo
http://doi.org/10.5281/zenodo.3735723

# Supplementary Material: Simultaneous two-color snapshot view on ultrafast charge and spin dynamics in a Fe-Cu-Ni tri-layer

Benedikt Rösner,[1*] Boris Vodungbo,[2] Valentin Chardonnet,[2] Florian Döring,[1] Vitaliy A. Guzenko,[1] Marcel Hennes,[2] Armin Kleibert,[1] Maxime Lebugle,[1] Jan Lüning,[2] Nicola Mahne,[3] Aladine Merhe,[2] Denys Naumenko,[4] Ivaylo P. Nikolov,[4] Ignacio Lopez-Quintas,[4] Emanuele Pedersoli,[4] Primož R. Ribič,[4,5] Tatiana Savchenko,[1] Benjamin Watts,[1] Marco Zangrando,[3,4] Flavio Capotondi,[4] Christian David[1], Emmanuelle Jal[2]

[1] Paul Scherrer Institut, 5232 Villigen PSI, Switzerland

[2] Sorbonne Université, CNRS, Laboratoire de Chimie Physique – Matière et Rayonnement, LCPMR, 75005 Paris, France

[3] IOM-CNR, Strada Statale 14-km 163,5, 34149 Basovizza, Trieste, Italy

[4] Elettra-Sincrotrone Trieste, Strada Statale 14-km 163,5, 34149 Basovizza, Trieste, Italy

[5] Laboratory of Quantum Optics, University of Nova Gorica, 5001 Nova Gorica, Slovenia

## EXPERIMENTAL GEOMETRY

The experiment was performed in the DiProI Ultra High Vacuum chamber of FERMI [F. Capotondi *et al.* Journal of Synchrotron Radiation, 22 - 3, 544-552 (2015)]. As represented in Figure S1a, we used an x-y-z stage to position the twin-focus, off-axis zone plate, a sample x-y-z stage tilted for a 30° angle between the sample surface normal and the pre-zone plate beam propagation trajectory. With the set of parameters for the off-axis zone plates with an area of 3.8 mm x 3.8 mm, the resulting focal length is 74.4 mm and the spot displacement from the optical axis is 2.3 mm. A set of permanent magnets were set up around the sample to apply a field in the sample plane up to 130 mT in both directions. A CCD camera was placed approx. 220 mm downstream of the sample (resulting in a magnification of the projection of three) and oriented for normal incidence of the 1st order beam from the twin-focus, off-axis zone plate in order to image the beam projection transmitted through the sample. Finally, the infrared (IR) pump laser beam was overlapped with the "sample" 1st order focus position of the twin-focus, off-axis zone plate on the sample tri-layer magnetic film with an angle of 10° with respect to the average beam trajectory of the zone plates 1st order focus. From the difference of the sample tilt and the diffraction angle of the XUV beam at the zone plate, we calculate an XUV incidence angle at the sample of 18° from the sample surface normal.

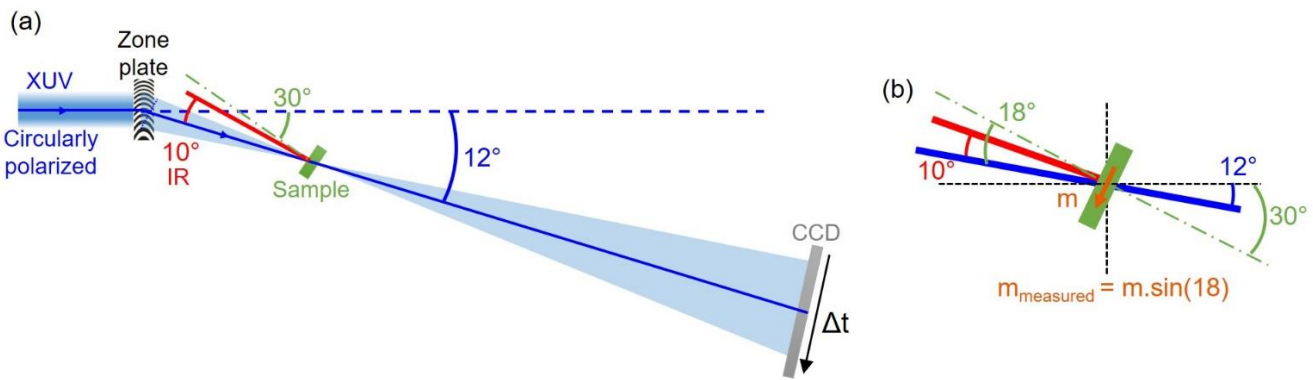

**Figure S1 | Schematic illustration of the experimental geometry. a,** Experimental scheme with angles between the sample surface normal and the incident XUV beam (green), between 1st zone plate diffraction order and incident XUV (blue), and between 1st zone plate diffraction order and incident IR pump (red). Note that the twin foci (pumped sample and unpumped reference spots, compare Figure 2 in main paper) are superimposed from this perspective. **b,** Zoom at the sample plane with representation of the in-plane magnetization of the metal tri-layer.

Thus, the apparent magnetization is equal to the real magnetization multiplied by sin(18°) and the IR pump arrives with an incidence angle of 8° (Figure S1b). This small crossing angle between IR and FEL radiation ensure a minimal broadening of the time resolution (about 1%) with respect to the ideal collinear geometry between FEL and IR pulses, since the stretching of the IR pulse on the sample surface is proportional to the inverse of cosine of the crossing angle.

**NORMALIZING DATA AND RETRIEVING XAS AND XMCD ABSORPTION CURVES**

As discussed in the main text, the twin-focus, off-axis zone plate projects two images onto the detector that correspond to photons coming from the two focus spots that illuminate the pumped sample area and unpumped reference area of the tri-layer magnetic film. Figure S2 shows the raw and normalized data accumulated with 600 FEL shots, for one applied magnetic field direction, and right-circularly polarized incoming XUV. Figure S2a shows the images recorded while pumping the sample at focus position with IR (sample area in the upper image; reference area shown in the lower image), while Figure S2b shows the respective images recorded without an IR pump pulse.

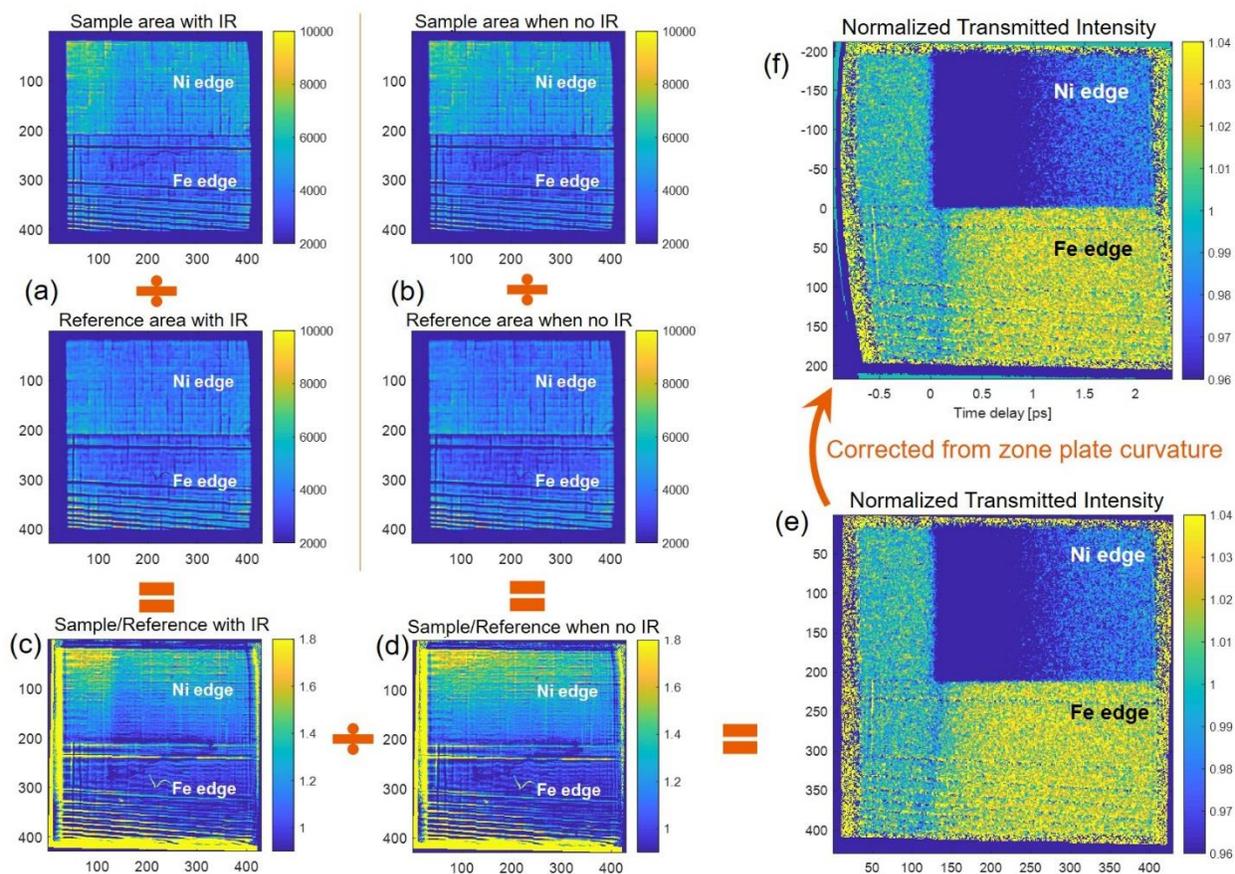

**Figure S2 | Transmitted intensity of circularly right polarized XUV through the tri-layer Fe(10nm)/Cu(2nm)/Ni(10nm) for one applied field direction. a,** Case for the two spots on the sample that are the pumped and reference areas with the IR pump switched on and **b,** for both spots without IR. Division of pumped area by the reference one for IR on **(c)** and IR off **(d)**. **e,** Division of Figure (c) and (d) to get the normalized transmitted intensity as explained in the text. **f,** Compensation of the zone plate curvature and an additional (linear) time streaking induced by the beam splitting. This is done by mapping the time information pixel by pixel calculating the path length to the focus spot from the respective area on the zone plate.

In general, the intensity recorded by the detector, $I$, can be broken down into contributions from the FEL source, $I_0$, the optics and detector efficiency, $\eta$, and the interaction with the sample that follows the Beer-Lambert law [A. Beer, Annalen der Physik und Chemie 86, 78-88 (1852)]. In the case where the detector images have been corrected for tilts and curvatures, information about the time, $t$, and photon energy, $E$, are encoded into the $x$ and $y$ pixel-positions of the recorded images, respectively, to give:

$$I(S, x, y) = I_0(S, E(y))\, \eta(E(y), x, y)\, e^{-\mu(E(y), t(x)) \cdot z} \quad \textbf{(eq. 1)}$$

where $S$ is the FEL shot index identifying the X-ray probe pulse, $\mu$ is the linear photoabsorption coefficient of the sample film and $z$ its thickness. Each detector image contains regions corresponding to the sample and reference focus spot on the sample. Dividing these without providing an IR pump pulse gives the relative optics and detector efficiencies of the corresponding regions:

$$\begin{aligned}\frac{I_s(S_i, x_s, y_s)}{I_r(S_i, x_r, y_r)} &= \frac{I_0(S_i, E(y))\, \eta(E(y), x_s, y_s)\, e^{-\mu_s(E(y)) \cdot z}}{I_0(S_i, E(y))\, \eta(E(y), x_r, y_r)\, e^{-\mu_r(E(y)) \cdot z}} \\ &= \frac{\eta(E(y), x_s, y_s)}{\eta(E(y), x_r, y_r)} e^{-[\mu_s(E(y)) - \mu_r(E(y))] \cdot z}\end{aligned} \quad \textbf{(eq. 2)}$$

where the $s$ and $r$ subscripts refer to the sample and reference detector image regions. To superimpose the two detector image areas, we used the Matlab affine2d function which mainly translates the image (besides a slight scaling to correct camera alignment errors) by a two-dimensional geometric transformation, [https://mathworks.com/help/images/ref/affine2d.html]. Note that simultaneous measurement of the sample and reference positions allows equation 2 to cancel the shot-to-shot XUV intensity variations without an extra $I_0$ measurement, which can be challenging at Free Electron Lasers [Higley et al., Rev. Sci. Instrum. 87, 033110 (2016)].

Furthermore, $\mu$ does not vary as a function of time without any IR pumping and strong similarity between the magnetic film in the sample and reference focus positions should make the exponential term close to unity. (*i.e.* $\mu_s = \mu_r$). The relative efficiency term is then expected to dominate the structures observed in this image ratio, which is shown in Figure S2d and appears to chiefly consist of three types of imperfections in the optical system: spatially inhomogeneous intensity distribution of the incident XUV pulse due an imperfectly flat wave front, inhomogeneities in the XUV diffraction efficiency of the twin-focus, off-axis zone plate, and stitching errors during the lithography step in the fabrication process of the twin-focus off-axis zone plate.

If we now pump with IR the material at the sample focus spot (indicating affected terms by an asterisk), we get:

$$\begin{aligned}\frac{I_s^*(S_i, x_s, y_s)}{I_r(S_i, x_r, y_r)} &= \frac{I_0(S_i, E(y))\, \eta(E(y), x_s, y_s)\, e^{-\mu_s^*(E(y), t(x)) \cdot z}}{I_0(S_i, E(y))\, \eta(E(y), x_r, y_r)\, e^{-\mu_r(E(y)) \cdot z}} \\ &= \frac{\eta(E(y), x_s, y_s)}{\eta(E(y), x_r, y_r)} e^{-[\mu_s^*(E(y), t(x)) - \mu_r(E(y))] \cdot z}\end{aligned} \quad \textbf{(eq. 3)}$$

Equation 3 says that taking the ratio of the pumped sample and reference regions of the detector image provides an approximate measure of the change in sample absorption as a function of time and energy encoded in the $x$ and $y$ pixel-positions. The precision of this normalization depends on the assumptions that the combined efficiency of the twin-focus, off-axis zone plate and the detector is trivial (*i.e.* equal to

1) and that the absorption coefficient of the material at the *sample* and *reference* focus positions are identical in the absence of pumping (*i.e.* $\mu_s=\mu_r$). This normalization is analogous to the "clean monitor" normalization method for TEY NEXAFS measurements described by Watts et al. [Watts et al., J. Electron Spectrosc. Relat. Phenom. 151, 105-120 (2006)] and is sufficient to observe some details of the dynamics as shown in Figure S2c.

However, the approximating assumptions in equation 3 can be removed by combining a pair of measurements from separate FEL shots (the $i^{th}$ shot coinciding with a pump, the $j^{th}$ not pumped) by dividing equation 3 by equation 2:

$$\frac{I_s^*(S_i,x_s,y_s)}{I_r(S_i,x_r,y_r)} \cdot \frac{I_r(S_j,x_r,y_r)}{I_s(S_j,x_s,y_s)} = e^{-[\mu_s^*(E(y),t(x))-\mu_s(E(y))]\cdot z}$$

$$= e^{-\Delta\mu_s(E(y),t(x))\cdot z}$$

(eq. 4)

This normalization is analogous to the "stable monitor" NEXAFS normalization method [Watts et al., J. Electron Spectrosc. Relat. Phenom. 151, 105-120 (2006)] that is based on the more reliable assumption that the relative efficiency and reference absorption terms ($\eta$ and $\mu_r$) are constant between shots. Figure S2e demonstrates that normalizing the data by the "stable monitor" method has removed the strong, nearly horizontal streaks from the image, flattening the left-hand, pre-IR-pump section of the image and bringing the system dynamics contrast well above the distortion and noise. A final correction of the image distortion caused by the curvature of the zone plate is applied by mapping each pixel to its respective time delay according to the local zone number [Buzzi et al., Sci. Rep. 7, 7253 (2017)] and the additional contribution to the path length by the beam splitting, see in Figure S2f and Figure 3b in the main text. From this figure and equation 4, we just need to integrate the two signals along the y-axis (the top part for Ni and the bottom part for Fe), and to perform the negative logarithm on this transmitted intensity ratio and divide by the corresponding sample layer thickness, *z*, to obtain the relative absorption as a function of the time delay, $\Delta\mu(t)$. Measuring the normalized transmitted intensity for the two directions of the magnetic field gives information on the two relative magnetic absorption spectra $\Delta\mu^{+/-}(t)$.

In order to compare relative absorption for different element and probing energies, one need to divide $\Delta\mu^{+/-}(t)$ by the static absorption $\mu_0^{+/-}$. In our case, we do not have an absolute measurement of the incident XUV intensity $I_0$, and therefore it is not possible to extract directly $\mu_0^{+/-}$. However, since we are measuring a reference area, with the theoretical transmission of the sample we can retrieve $I_0$, and therefore $\mu_0^{+/-}$. Indeed, $I = T \cdot I_0$ where $T$ is the total transmission that can be calculated by multiplying the transmission of each layer composing the sample. To calculate the transmission of each layer we have used the CXRO website [https://henke.lbl.gov/optical_constants/filter2.html] and the optical constant recently measured by Willems *et al.* [Willems et al., Phys. Rev. Letter 122, 217202 (2016)] for Fe and Ni at resonance for both magnetic field directions. We obtained a total transmission for an XUV probe of 65.8 eV (Ni edge) of 0.112 and 0.116 for a positive and negative magnetic field respectively, and for an XUV probe of 55.7 eV (Fe edge) of 0.062 and 0.069 for a positive and negative magnetic field respectively. Note that since we are close to normal incidence, the reflectivity at each interfaces is negligible. With this theoretical transmission we retrieve $\mu_0^{+/-}$, and therefore look at relative absorption $\Delta\mu^{+/-}(t)/\mu_0^{+/-}$ as shown in Figure 4b of the main text. In the Figure 4c, the error bars arise from the average of the curves measured with different experimental configuration (reversing the applied magnetic field or the circular XUV polarization).

From the relative magnetic absorption spectra $\Delta\mu^{+/-}(t)$, the pure relative absorption can be retrieved, which is sensitive solely to the electronic charge, $\Delta XAS(t) = \frac{\Delta\mu^+ + \Delta\mu^-}{2}$, and the relative XMCD signal, which is sensitive to the electronic spin, $\Delta XMCD(t) = \Delta\mu^+ - \Delta\mu^-$. We do the same for the static absorption, $XAS_0 = \frac{\mu_0^+ + \mu_0^-}{2}$ and magnetization, $XMCD_0 = \mu_0^+ - \mu_0^-$, and can extract the normalized relative absorption $\Delta XAS/XAS_0$ and $\Delta XMCD/XMCD_0$ as shown in Figure 4d and 4e in the main text.

## FITTING PARAMETERS

To extract the time constant given in the main text, we fitted the $\Delta XMCD$ data by the equation derived by Malinowski et al. [Malinowski et al., Nat. Phys. 4, 855 (2008)]:

$$\Delta XMCD(t) = \left[\frac{-b}{\sqrt{t/d+1}} + \frac{ct_1 - bt_2}{t_2 - t_1}e^{-t/t_1} + \frac{t_2(b-c)}{t_2 - t_1}e^{-t/t_2}\right]\Theta(t) \otimes \Gamma(t), \quad \textbf{(eq. 5)}$$

where $b$ is the amplitude after equilibrium, $c$ is the initial electron temperature rise, $d$ is the cooling by heat diffusion time (fixed to 50 ps), $t_1$ is the demagnetization time and $t_2$ is the recovery time. The $\Theta$ function is the Heaviside function, which we are convoluting by $\Gamma$, a Gaussian function with 120 fs full width at half maximum (FWHM) that accounts for the temporal resolution degradation corresponding to the mean squared duration of the XUV and IR pulses of 70 and 100 fs, respectively. Note that even if the resolution of the response is depending on the quadrature of the width of pump and probe pulses, the resolution of the experiment is of 2.5 fs by construction.

For the fitting of $\Delta XAS$, we systematically screened constants b and c in equation 5. The parameters giving the fit shown in Figure 4d of the main text are summarized in the table below.

**Table 1 | Fit parameters used in equation 5.**

|  | b | c | $t_1$ (fs) | $t_2$ (ps) |
|---|---|---|---|---|
| D-XAS Fe | 0.005 | -0.007 | 20 | 0.2 |
| D-XAS Ni | -0.005 | -0.07 | 60 | 0.8 |
| D-XMCD Fe | 0 | 0.24 | 450 | 10 |
| D-XMCD Ni | 0 | 0.35 | 250 | 2 |

## ABSORPTION PROFILE

To quantify the IR absorption profile in our sample, we have calculated the attenuation of IR radiation using the transfer matrix method [https://arxiv.org/abs/1603.02720] and the "tmm" python software package [http://pypi.python.org/pypi/tmm]. The optical constants are derived from the standard values provided by Palik [Palik, Handbook of Optical Constants of Solids, Academic Press 1997], listed in the table of Figure S3. The local absorption coefficient and integrated absorption through the different layers are shown in Figure S3.

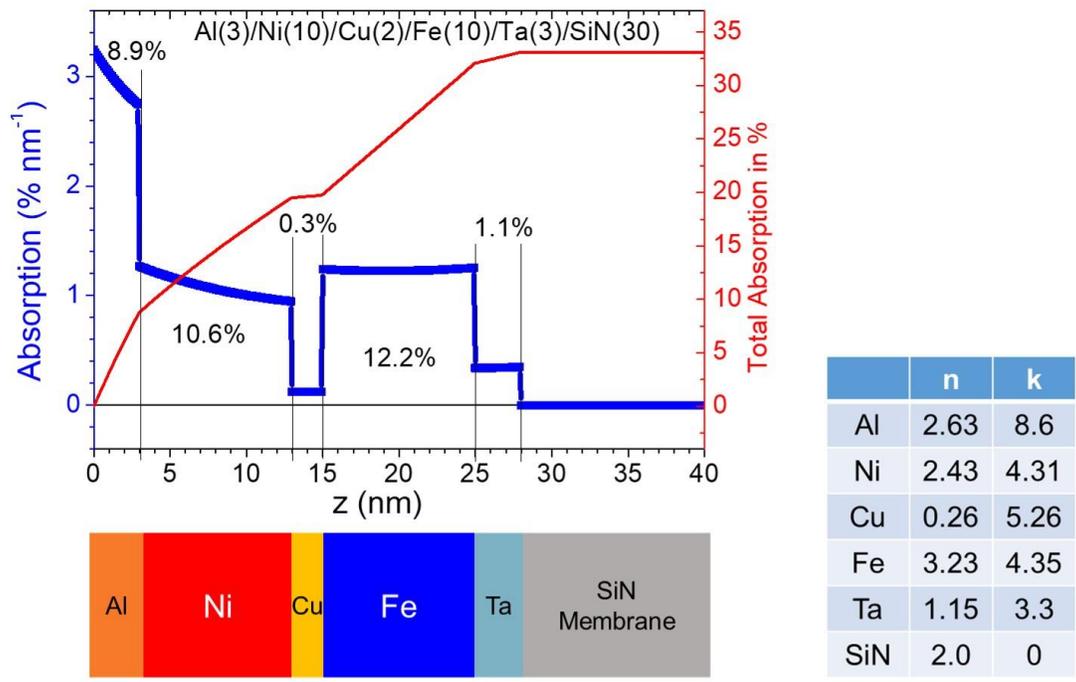

**Figure S3 | IR absorption profile by our tri-layer samples.** (blue) absorption rate in percentage per nanometer. (red) integration of the blue curve to demonstrate the total absorption at each layer (integral values for individual layers labelled on the graph in black).

20